\pdfoutput=1

\documentclass[sigconf,screen,authorversion]{acmart} 

\usepackage[utf8]{inputenc}        
\usepackage{csquotes}              
\usepackage[nameinlink]{cleveref}  
\usepackage{pgfplots}              
\usepackage{filecontents}          
\usepackage{silence}               

\usetikzlibrary{patterns} 
\pgfplotsset{compat=1.14} 
\copyrightyear{2020}
\acmYear{2020}
\setcopyright{acmlicensed}
\acmConference[SIGCSE '20]{The 51st ACM Technical Symposium on Computer Science Education}{March 11--14, 2020}{Portland, OR, USA}
\acmBooktitle{The 51st ACM Technical Symposium on Computer Science Education (SIGCSE '20), March 11--14, 2020, Portland, OR, USA}
\acmPrice{15.00}
\acmDOI{10.1145/3328778.3366816}
\acmISBN{978-1-4503-6793-6/20/03}

\begin{document}
\fancyhead{}

\title{What Are Cybersecurity Education Papers About? A Systematic Literature Review of SIGCSE and ITiCSE Conferences}

\author{Valdemar Švábenský}
\orcid{0000-0001-8546-280X}
\affiliation{
  \institution{Masaryk University}
  \country{Czech Republic}
}
\email{svabensky@ics.muni.cz}

\author{Jan Vykopal}
\orcid{0000-0002-3425-0951}
\affiliation{
  \institution{Masaryk University}
  \country{Czech Republic}
}
\email{vykopal@ics.muni.cz}

\author{Pavel Čeleda}
\orcid{0000-0002-3338-2856}
\affiliation{
  \institution{Masaryk University}
  \country{Czech Republic}
}
\email{celeda@ics.muni.cz}

\begin{abstract}
Cybersecurity is now more important than ever, and so is education in this field. However, the cybersecurity domain encompasses an extensive set of concepts, which can be taught in different ways and contexts. To understand the state of the art of cybersecurity education and related research, we examine papers from the ACM SIGCSE and ACM ITiCSE conferences. From 2010 to 2019, a total of 1,748 papers were published at these conferences, and 71 of them focus on cybersecurity education. The papers discuss courses, tools, exercises, and teaching approaches. For each paper, we map the covered topics, teaching context, evaluation methods, impact, and the community of authors. We discovered that the technical topic areas are evenly covered (the most prominent being secure programming, network security, and offensive security), and human aspects, such as privacy and social engineering, are present as well. The interventions described in SIGCSE and ITiCSE papers predominantly focus on tertiary education in the USA. The subsequent evaluation mostly consists of collecting students' subjective perceptions via questionnaires. However, less than a third of the papers provide supplementary materials for other educators, and none of the authors published their dataset. Our results provide orientation in the area, a synthesis of trends, and implications for further research. Therefore, they are relevant for instructors, researchers, and anyone new in the field of cybersecurity education. The information we collected and synthesized from individual papers are organized in a publicly available dataset.
\end{abstract}



\maketitle

\section{Introduction}
\label{sec:intro}

With the rising importance of cybersecurity and its global job vacancy increasing to 2.93 million~\cite{isc2}, there is an imminent need for training more cybersecurity workers. As a result, a broad range of educational initiatives arose to address this need. In 2013, cybersecurity was included in ACM/IEEE computing curricula~\cite{curricula2013}. Four years later, the Joint Task Force on Cybersecurity Education (JTF) published comprehensive curricular guidance~\cite{cybered} to help educators design cybersecurity courses. In addition, informal methods of education, such as extracurricular events~\cite{Dunn:2018} and competitions~\cite{taylor2017} for learners of all ages and expertise are gaining popularity.

SIGCSE 2020 begins a new decade of computing education research. Therefore, we feel it is appropriate to review the research advancements and teaching methods we have seen over the last ten years. In this paper, we examine the development and state of the art of cybersecurity education as presented at ACM SIGCSE and ACM ITiCSE conferences from 2010 to 2019. We have chosen these two conferences because they represent the leading venues in the area of computing education. Apart from their rich history and a large number of quality submissions, they are currently the only two conferences in the field that rank as CORE A~\cite{core}.

This literature review brings several contributions to various target groups. For cybersecurity instructors and educational managers, it shows what topics are taught and how. For researchers, it provides an overview of evaluation methods, implications for further research, and practical recommendations. Finally, for the SIGCSE/ITiCSE community as a whole, it serves as a snapshot of ten years of the latest development and synthesis of accepted papers.

\begin{table*}
  \setlength\tabcolsep{1pt}
  \caption{Number of full papers at SIGCSE | ITiCSE conferences over the years 2010--2019 after each step of the literature review}
  \label{tab:num-papers-development}
  \begin{tabular}{lp{2.4mm}rcrp{2.8mm}rcrp{2.8mm}rcrp{2.8mm}rcrp{2.8mm}rcrp{2.8mm}rcrp{2.8mm}rcrp{2.8mm}rcrp{2.8mm}rcrp{2.8mm}rcrp{2.8mm}rcr}
\toprule
SIGCSE | ITiCSE & & \multicolumn{3}{c}{2010} & & \multicolumn{3}{c}{2011} & & \multicolumn{3}{c}{2012} & & \multicolumn{3}{c}{2013} & & \multicolumn{3}{c}{2014} & & \multicolumn{3}{c}{2015} & & \multicolumn{3}{c}{2016} & & \multicolumn{3}{c}{2017} & & \multicolumn{3}{c}{2018} & & \multicolumn{3}{c}{2019} & & \multicolumn{3}{c}{\textbf{Total}} \\
\midrule
Published papers & &
103 & | & 60 & & 
107 & | & 66 & & 
100 & | & 61 & & 
111 & | & 51 & &
108 & | & 53 & &
105 & | & 54 & &
105 & | & 51 & &
105 & | & 56 & &
161 & | & 56 & &
169 & | & 66 & &
\textbf{1174} & | & \textbf{574} \\
Candidate papers & & 
5 & | & 2 & & 
4 & | & 1 & & 
1 & | & 2 & &
8 & | & 2 & &
3 & | & 2 & &
6 & | & 2 & &
7 & | & 3 & &
5 & | & 1 & &
14 & | & 4 & &
15 & | & 3 & &
\textbf{68} & | & \textbf{22} \\
Selected papers & &
5 & | & 1 & & 
3 & | & 1 & & 
0 & | & 2 & &
6 & | & 1 & &
2 & | & 2 & &
5 & | & 2 & &
6 & | & 1 & &
5 & | & 1 & &
12 & | & 3 & &
11 & | & 2 & &
\textbf{55} & | & \textbf{16} \\
\bottomrule
\end{tabular}
\end{table*}

\section{Related Work}

This section maps related primary and secondary studies in two areas: cybersecurity education and computing education in general. We build upon and extend the below-mentioned works by focusing on the cybersecurity domain at SIGCSE and ITiCSE.

\subsection{Cybersecurity Education}

Although some papers reviewed or synthesized results of cybersecurity education efforts, none of them focused on SIGCSE and ITiCSE. Fujs et al.~\cite{Fujs:2019} performed a literature review on using qualitative methods in cybersecurity research, which includes research on cybersecurity education. Next, Cabaj et al.~\cite{cabaj2018} examined 21 cybersecurity Master degree programs and their mapping to ACM/IEEE curricular guidelines~\cite{curricula2013}. Jones et al.~\cite{Jones:2018} studied the core knowledge, skills, and abilities that cybersecurity professionals need. Parrish et al.~\cite{Parrish:2018} provided an overview of cybersecurity curricula.

Also, there are synthesis papers on the topic of Capture the Flag (CTF). This popular format of competitions or educational events allows the participants to practice their cybersecurity skills. Taylor et al.~\cite{taylor2017} mapped the state of the art of 36 implementations of CTF. Next, Burns et al.~\cite{burns2017} analyzed the solutions of 3,600 CTF challenges from 160 competitions to discover which challenge types are the most popular and what is their difficulty. Finally, Vigna et al.~\cite{vigna2014} presented a framework for organizing CTF events, building upon ten years of experience with running the largest educational CTF.

\subsection{Other Areas of Computing Education}

Many literature surveys in computing education focus on programming. ITiCSE Working Groups often create comprehensive reviews, such as the one by Luxton-Reilly et al.~\cite{Luxton-Reilly:2018} on introductory programming education. The authors inspected 1,666 papers from 15 years. Ihantola et al.~\cite{ihantola2015} surveyed a decade of 76 papers on educational data mining and learning analytics in programming. Next, Becker and Quille~\cite{Becker:2019} reviewed topics of 481 papers about CS1 from 50 years of SIGCSE conferences. Finally, Keuning et al.~\cite{Keuning:2018} compared 101 tools that provide automated feedback to students who solve programming assignments. All these studies provide an excellent example of conducting literature reviews.

\section{Method of Conducting the Review}  

There are two similar approaches for surveying research papers: a systematic literature review (SLR) and a systematic mapping study (SMS). Both approaches aim at summarizing the existing research, discovering research trends, and identifying gaps in current research. Although SLR and SMS slightly differ (see~\cite{kitchenham2007, Petersen:2008, petersen2015}), these differences are negligible for the purposes of this paper, which could be classified as either SLR or SMS.
The methods we applied follow the well-established guidelines for SLR~\cite{kitchenham2007} and SMS~\cite{Petersen:2008, petersen2015}.

\subsection{Research Questions}
\label{subsec:rqs}

Our literature review examines five research questions:
\begin{enumerate}
    \item \textit{What cybersecurity topics are discussed in the papers?}
    \item \textit{How and in what context are the topics taught?}
    \item \textit{How are the teaching interventions evaluated?}
    \item \textit{What is the impact of the published papers?}
    \item \textit{Who are the members of the cybersecurity education community at SIGCSE and ITiCSE?}
\end{enumerate}
Our motivation for each of the research questions was:
\begin{enumerate}
    \item To discover which topics are taught and researched a lot and whether there are any underrepresented topics.
    \item To describe the common teaching practices.
    \item To examine research methods in cybersecurity education.
    \item To find out whether the community adopts the published teaching innovations and research results.
    \item To understand who forms the SIGCSE/ITiCSE community.
\end{enumerate}
We address the research questions by extracting and analyzing data from 71 papers. The following Sections 3.2--3.4 describe the process. \Cref{tab:num-papers-development} provides context by showing how many papers were published within the selected time range. It also presents how the number of papers evolved at each step of our literature review.

\subsection{Paper Search and Automated Filtering}

We searched for papers in the ACM Digital Library by submitting the query: \texttt{cybersecur* OR secur*}. We used this broad search term to cover a wide area and avoid the risk of missing a relevant paper. We then refined the results to publications in conference proceedings since 2010. Further restriction on the SIGCSE conference yielded 209 results; the ITiCSE conference yielded 52 results. We searched for SIGCSE papers on May 30, 2019, and for ITiCSE papers on July 15, 2019. We also searched for ICER conference papers (on August 15, 2019), which surprisingly yielded 0 results.

As the next step, we removed 1- or 2-pages long submissions, which included, for example, poster abstracts and panel proposals. Although these submissions indicate a general topic of interest, they do not have enough space to discuss details relevant to our research questions. Afterward, we were left with 90 candidate full papers (68 from SIGCSE and 22 from ITiCSE), each 5--7 pages long.

\subsection{Pilot Reading and Manual Filtering}

Two authors preliminarily read the 90 papers to mark candidate false positives. The reading was independent to prevent bias. The resulting inter-rater agreement measured by Cohen's kappa~\cite{landis1977} was 0.87, which is an \enquote{almost perfect agreement}. The remaining discrepancies were only minor and were resolved by discussion. In the end, we eliminated 19 false positives, selecting 71 papers for the literature review. The most common reason for exclusion was that the paper mentioned the word \enquote{security}, for example, in the general context that it is an essential concept, but did not deal with anything related to cybersecurity education.

\subsection{Full Text Reading and Data Extraction}

For each research question, we drafted several criteria that defined what kind of data we would extract from the papers. Subsequently, we performed a pilot test on 30 papers to see if these criteria are reasonable and if the data can indeed be extracted. Over the course of this action and several discussions among the authors, the criteria iteratively evolved. We present their final form in \Cref{sec:results} along with the corresponding result to save space and improve readability.

Upon agreeing on the data extraction criteria and process, we thoroughly read the selected 71 papers and documented the results, creating our dataset~\cite{dataset}. When in doubt, we discussed unclear cases to reach agreement. \Cref{sec:results} presents the synthesized results.

\subsection{Limitations}

This review is limited by narrowing its scope only to the SIGCSE, ITiCSE, and ICER conferences. However, other conferences, such as IEEE FIE and USENIX ASE/3GSE, also publish cybersecurity education papers. There are related journals as well, such as ACM TOCE, ACM Inroads, and Elsevier Computers and Education. Nevertheless, as mentioned in \Cref{sec:intro}, the SIGCSE and ITiCSE conferences are the flagships in the field, and so we consider them representative of the trends within the cybersecurity education community.

We are confident that our broad search query captured all relevant papers. The filtering of false positives was double-checked. However, the data extraction from the papers might be problematic. Since it was done manually, the readers may have overlooked or misinterpreted something. Nevertheless, we performed cross-author discussion and validation to minimize the risk of incorrectness.

\section{Results}
\label{sec:results}

We now present the selected results that synthesize the data extracted from the 71 papers. The dataset of full records is publicly available~\cite{dataset}. Each Section 4.1--4.5 is mapped to the corresponding research question from \Cref{subsec:rqs}. When needed, we include examples of representative papers within the category\footnote{A single paper may correspond to multiple categories. Therefore, the counts reported in respective categories sometimes add up to more than the total of 71 papers. Also note that some data, such as the citation counts or the availability of hyperlinks, are bound to the time of writing this section (the second half of August 2019).}.

\subsection{RQ1: What Topics Are Taught?}

To address the first research question, we identified general topic areas as listed in the JTF Cybersecurity Curriculum~\cite{cybered}, as well as specific topics discussed in the papers.

\subsubsection{Which Knowledge Areas from the JTF Cybersecurity Curriculum do the papers focus on?}

\begin{figure}[t]
  \centering
  \begin{tikzpicture}
  \begin{axis}[
    xbar,
    xmin=0,
    width=6.4cm, height=6.2cm, enlarge y limits=0.1,
    xlabel={Number of papers},
    symbolic y coords={8 Societal security, 7 Organizational security, 6 Human security, 5 System security, 4 Connection security, 3 Component security, 2 Software security, 1 Data security},
    ytick=data,
    ytick style={opacity=0},
    nodes near coords, nodes near coords align={horizontal},
    ]
    \addplot[black,pattern=north east lines] coordinates {(8,8 Societal security) (7,7 Organizational security) (13,6 Human security) (18,5 System security) (20,4 Connection security) (4,3 Component security) (20,2 Software security) (29,1 Data security)};
  \end{axis}
\end{tikzpicture}
  \vspace*{-0.5cm}
  \caption{The distribution of how often each of the eight JTF Cybersecurity Curriculum Knowledge Areas~\cite{cybered} was discussed in the selected papers.}
  \label{fig:topics}
\end{figure}
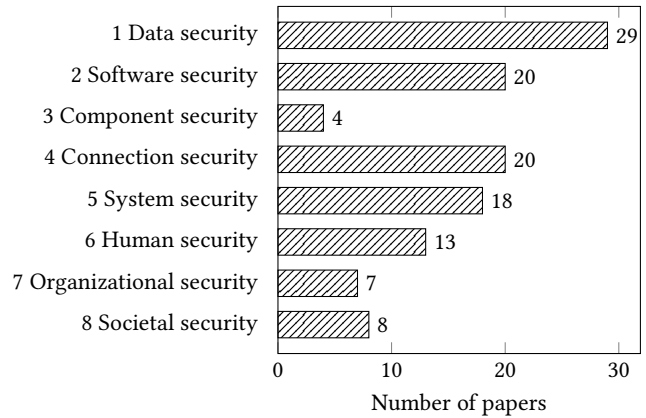

The curriculum~\cite{cybered} consists of eight Knowledge Areas (KA), each including several Knowledge Units (KU). \Cref{fig:topics} shows the distribution of how often each KA was present. The most frequent KUs (in 10 or more papers) were Network Defense (19 papers), Implementation (18 papers), System Control (15 papers), and Cryptography (14 papers). Nevertheless, performing this mapping was sometimes tricky. The papers differ in the level of detail when describing the teaching content, and only 10 papers reference a standardized curriculum to define their learning objectives (most frequently ACM/IEEE guidelines~\cite{curricula2013}).

\subsubsection{What are the primary cybersecurity topics?}
We performed open coding of the topics that the individual papers focused on. The most commonly assigned codes (to 10 or more papers) were:
\begin{itemize}
    \item Secure programming and software development, including reverse engineering (24 papers).
    \item Network security and monitoring (23 papers).
    \item Cyber attacks, malware, hacking, offensive security, and exploitation (17 papers).
    \item Human aspects, including privacy, social engineering, law, ethics, and societal impact (17 papers).
    \item Cryptography (15 papers).
    \item Authentication and authorization (13 papers).
\end{itemize}
While this coding was performed independently from the mapping to the JTF Curriculum, the findings are similar to the most frequent KUs mentioned earlier, supporting the validity of the results.

\subsection{RQ2: How Are the Topics Taught?}

Next, we examined the target groups, teaching methods, time frames, and student population sizes.

\subsubsection{Whom are the teaching interventions aimed at?}
The most prominent target group for teaching interventions in 54 papers are university or college students (undergraduates or graduates\footnote{Including midshipmen or cadets in military academies in 3 papers.}). Other target groups are in the minority: instructors and educational managers in 7 papers, K-12 students (middle or high school) in 7 papers, and professional learners in 3 papers. Lastly, 8 papers do not explicitly report their target group, although in most cases, the paper content suggests university students.

\subsubsection{Which teaching methods are used?}
The subject of 64 papers (90\%) is something that was applied in teaching practice, such as a tool, an exercise, or a full course. The remaining 7 papers only propose or discuss an idea without having it used in actual teaching. Within these 64 papers that describe a teaching intervention, the most common teaching method mentioned in 51 papers is some form of hands-on learning during class time or self-study. This includes labs, exercises, practical assignments, educational games, and other activities for practicing the selected topic. Other common teaching methods are lectures (24 papers), long-term projects that typically span the whole semester (10 papers), discussion (8 papers), and writing (7 papers). Exactly half of the 64 teaching papers mentions involving students in pairwork or groupwork.

\subsubsection{What is the time frame?}
4 papers study an intervention only lasting up to one hour, typically a single short assignment. Next, 8 papers deal with the time frame of several hours: typically an exercise or a workshop within a single day. 4 papers study a period of several days up to one week. 9 papers are within the time window of several weeks or one month. The most papers -- 23 -- study an intervention lasting more than 1 month, typically a semester-long course. Related to that, 27 papers describe experience or data from a single run of the intervention, 12 papers from two runs, and 17 papers from three or more runs.

\subsubsection{How big are the participant populations?}
\label{subsubsec:population}
Out of the 64 papers that discuss practical teaching interventions, we looked at how many students participated in it. Since the reporting was largely heterogeneous, we performed several simplifications. First, if the paper reported a repeated intervention, we considered the total sum of the participants. Second, if the paper reported a range, we considered the average number. Third, if the paper reported only the sample size present in the evaluation, we picked this number as the lower bound, although the real number of participants could have been higher. So, we report only what was discernible from the paper. The median number of participants was 62.5, and the distribution was as follows: 3 papers had 1--19 participants (minimum 17), 19 papers had 20--49 participants, 15 papers had 50--99 participants, and 17 papers had 100 or more participants (maximum 14,000). 9 papers did not report the number, and for~\cite{Chukuka:2016}, this was not applicable, as it dealt with ethical policies, not human participants.

\subsection{RQ3: How Is Evaluation Performed?}

This section examines whether the papers present any evaluation. If they do, we investigate what was evaluated, what data were collected, and how the data were analyzed. We finish the section by looking at sample sizes and publishing of datasets.

\subsubsection{What is the goal of the evaluation?}
We examined what aspects related to teaching and learning were evaluated in the papers. By doing so, we synthesized four evaluation goals (EG):
\begin{itemize}
    \item[EG1] \textit{Subjective perception of participants}, which includes learners' or teachers' attitudes and opinions of the subject of the paper (for example, a teaching intervention, course, or a cybersecurity exercise). The evaluation focuses on whether the participants reported self-perceived learning or found the subject useful, effective, understandable, easy/difficult, enjoyable, motivational, or supporting their career interest. There were 50 papers in this category, such as~\cite{Mack:2019}.
    \item[EG2] \textit{Objective learning of students}, which includes measuring learners' performance with summative assessment and computing test scores or grades to determine learning. There were 21 papers in this category, such as~\cite{Taylor:2011}.
    \item[EG3] \textit{Artifacts produced by students}, which includes examining submissions of assignments~\cite{Hooshangi:2015}, screen recordings from a learning environment~\cite{Tabassum:2018}, or logs from using a tool~\cite{Whitney:2015}. The focus was usually on better understanding students and their interactions, not necessarily measuring their learning as in EG2. There were 8 papers in this category.
    \item[EG4] \textit{Other artifacts}, which includes examining textbooks~\cite{Taylor:2019}, reviewing of teaching materials by teachers~\cite{George:2013}, or analyzing ethical policies~\cite{Chukuka:2016}. There were 5 papers in this category.
\end{itemize}
Finally, 8 papers did not present any evaluation, such as~\cite{Kazemi:2010} that only described a tool and its possible usage scenarios.

\subsubsection{What types of data are collected?}
\begin{figure}[t]
  \centering
  \begin{tikzpicture}

\draw (0.5,0) -- (4.5,0); 
\draw (0,0.5) -- (0,6.5); 

\foreach \x in {1,2,3,4} {\draw (\x,-2pt) -- (\x,2pt);}
\draw (1,-5pt) node[below] {Subj. (EG1)};
\draw (2,-18pt) node[below] {Obj. (EG2)};
\draw (3,-5pt) node[below] {Artif. (EG3)};
\draw (4,-18pt) node[below] {Other (EG4)};

\draw (-10pt,0) node[left] {No evaluation};
\draw (-2pt,1) node[left] {Questionnaires (DT1)} -- (2pt,1);
\draw (-2pt,2) node[left] {Tests (DT2)} -- (2pt,2);
\draw (-2pt,3) node[left] {Student artifacts (DT3)} -- (2pt,3);
\draw (-2pt,4) node[left] {Other artifacts (DT4)} -- (2pt,4);
\draw (-2pt,5) node[left] {Interviews (DT5)} -- (2pt,5);
\draw (-2pt,6) node[left] {Anecdotal only (DT6)} -- (2pt,6);

\draw (0,0) circle[radius=sqrt(8/20/pi)]  node {8};
\draw (1,1) circle[radius=sqrt(39/20/pi)] node {39};
\draw (1,5) circle[radius=sqrt(6/20/pi)]  node {6};
\draw (1,6) circle[radius=sqrt(7/20/pi)]  node {7};
\draw (2,2) circle[radius=sqrt(21/20/pi)] node {21};
\draw (3,2) circle[radius=sqrt(1/20/pi)]  node {1};
\draw (3,3) circle[radius=sqrt(8/20/pi)]  node {8};
\draw (3,5) circle[radius=sqrt(1/20/pi)]  node {1};
\draw (4,1) circle[radius=sqrt(1/20/pi)]  node {1};
\draw (4,4) circle[radius=sqrt(4/20/pi)]  node {4};

\end{tikzpicture}
  \vspace*{-0.2cm}
  \caption{The mapping of evaluation goals (EG) to the collected data types (DT) and the corresponding frequency.}
  \label{fig:evaluation}
\end{figure}
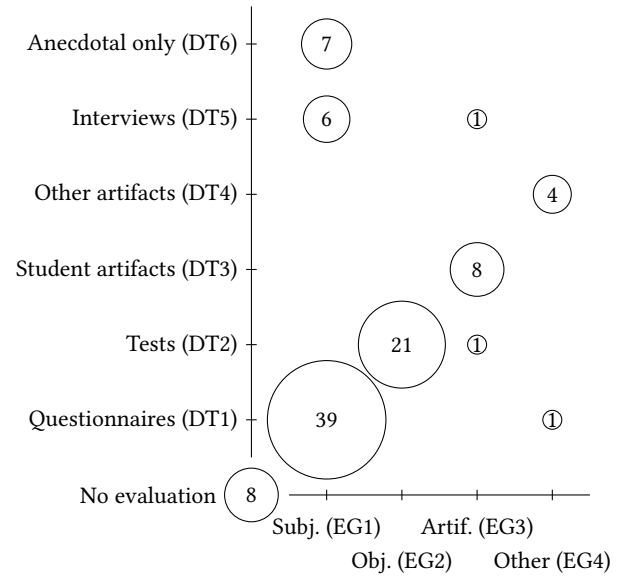

We synthesized six data types (DT) collected for the evaluation:
\begin{itemize}
    \item[DT1] \textit{Questionnaire data}, most often about participants' subjective perceptions (EG1). Out of 40 papers in this category, 26 employed post-intervention questionnaires only (such as~\cite{Timchenko:2015}), while the remaining 14 used pre- and post-questionnaire study design (such as~\cite{Egelman:2016}).
    \item[DT2] \textit{Test data}, the results of summative assessment, such as grades or the number of assignments solved, usually to measure objective learning (EG2). Out of 22 papers in this category, 10 employed post-test only (for example,~\cite{Mezher:2019}, although the test includes just two questions), and the remaining 12 used pre- and post-test study design (such as~\cite{Taylor:2011}).
    \item[DT3] \textit{Student artifacts} (see EG3 for examples). There were 8 papers in this category.
    \item[DT4] \textit{Other artifacts} (see EG4 for examples). There were 4 papers in this category.
    \item[DT5] \textit{Interviews} and focus group discussions to most often examine participants' subjective perceptions (EG1). There were 7 papers in this category, such as~\cite{Skirpan:2018}.
    \item[DT6] \textit{Anecdotal only}, if the paper did not present any of the above evidence types (DT1--DT5), but only reported the authors' or participants' experience, for example, from course feedback, not backed up with any research tool. There were 7 papers in this category, for example,~\cite{Salah:2014}.
\end{itemize}
\Cref{fig:evaluation} shows the mapping of EG to DT. The 8 papers that did not present any evaluation also did not collect any evidence.

\subsubsection{How are the collected data analyzed?}
We identified three analysis methods (AM) for interpreting the collected data:
\begin{itemize}
    \item[AM1] \textit{Descriptive statistics}, which were most often used to analyze questionnaire data (DT1) and test data (DT2). A majority of 54 papers employed this evaluation method, although the depth of the statistics varied from a single histogram~\cite{Xu:2012} to detailed tables including confidence intervals~\cite{Dunn:2018}.
    \item[AM2] \textit{Inferential statistics}, which involved formulating a hypothesis, applying an appropriate statistical test, and reporting the results, including p-value and sample size. There were 19 papers in this category, and although 4 more had attempted to use inferential statistics, the reporting had been insufficient. For example, although~\cite{Cappos:2014} presents a p-value, information about which test was applied is missing. Another transgression is present in~\cite{Jin:2018}: t-test was performed, but the statistic value and the p-value are not reported. On the contrary,~\cite{Egelman:2016} is a great example of a thorough statistical evaluation.
    \item[AM3] \textit{Qualitative analysis}, which included expert reviews of both student and non-student artifacts (DT3, DT4), their visual analysis, or thematic coding of interviews (DT5). There were 10 papers in this category.
\end{itemize}
15 papers did not employ any analysis method, since 8 of them did not collect any evidence, and the remaining 7 collected only anecdotal evidence, for which no special analysis was performed.

\subsubsection{What is the sample size?}
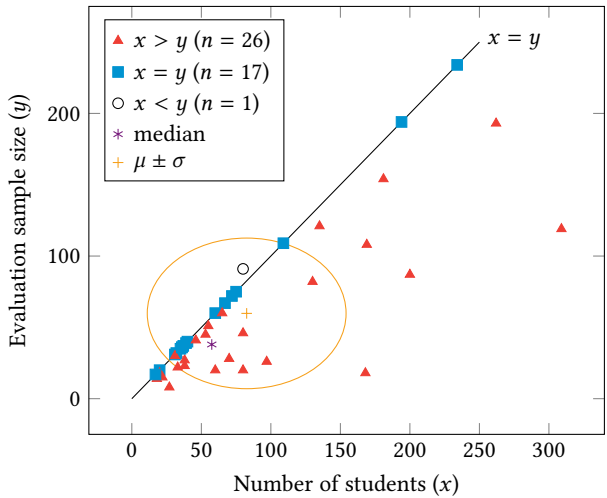
\begin{figure}[t]
  \centering


\begin{tikzpicture}
\begin{axis}[
legend pos=north west,
legend cell align=left,
xlabel=Number of students ($x$),
ylabel=Evaluation sample size ($y$)]

\addplot[
scatter/classes={
a={mark=triangle*,ACMRed},
b={mark=square*,ACMBlue},
c={mark=o,black},
d={mark=asterisk,ACMPurple},
e={mark=+,ACMOrange}
},
scatter,
only marks,
scatter src=explicit symbolic]
table[x=x,y=y,meta=label]{input.dat};

\addplot[
thin,
domain=0:250]
{x}
node[right,pos=1] {$x=y$};

\draw[color=ACMOrange] (82.6, 59.8) circle [x radius=71.5, y radius=52.8];

\legend{\:$x>y$ ($n=26$), \:$x=y$ ($n=17$), \:$x<y$ ($n=1$), \:median, \:$\mu\pm\sigma$}
\end{axis}
\end{tikzpicture}
  \caption{The number of students participating in a teaching intervention compared to the size of the sample that participated in the subsequent evaluation (44 data points).}
  \label{fig:sample-size}
\end{figure}

We also examined the sample size in papers that evaluated students. The median sample size was 40.5, and the distribution was as follows: 6 papers had 1--19 participants (minimum 8), 21 papers had 20--49 participants, 10 papers had 50--99 participants, and 11 papers had 100 or more participants (maximum 1,974). 11 papers did not report the number, and for 12 papers this was not applicable, as they either performed no evaluation or did not evaluate people. \Cref{fig:sample-size} puts this data\footnote{Two outliers with huge sample sizes were removed from the plot for readability. The remaining 25 papers either did not report $x$ or $y$ values, or this distinction was not applicable. The one point above the $x=y$ line is the paper~\cite{Mezher:2019}, in which the authors reported that 80 students participated in the presented exercise but then report evaluation results from a test taken by 91 students.} in context to the participant population reported in \Cref{subsubsec:population}. There is a visible trend that not all students participate in the evaluation.

\subsubsection{Can the evaluation be replicated?}
Surprisingly, none of the examined papers provides a link to the corresponding dataset that was analyzed. Although one paper~\cite{Deshpande:2019} includes a link to a Gitlab repository with the data, the repository is unavailable. Nevertheless, papers such as~\cite{Zinkus:2019} present exemplary practice by including the full wording of the questions that the evaluators asked.

\subsection{RQ4: What Is the Impact of the Papers?}

Our next question was whether the papers influenced other researchers and practitioners. However, as Malmi~\cite{Malmi:2015} argues, it is difficult to measure the impact of computing education research. He discusses counts of citations and paper downloads as two possible but imperfect metrics. For tools, he suggests download count as well, but this information is usually private and thus inapplicable. Therefore, we considered three metrics: providing a publicly available output, paper download count, and citation count.

\subsubsection{Do the papers provide output usable by other educators?}
30 of the 71 examined papers reference output for other educators. This most often includes course materials, lab modules/exercises, or software tools. The authors use dedicated websites, their institutional websites, or public repositories to share the content. However, out of the 30 linked websites, only 22 are still available, leaving 49 of the 71 papers (69\%) without any supplementary materials.

\subsubsection{How much are the papers downloaded and cited?}
The median cumulative number of downloads from the ACM Digital Library is 176 (min = 19, max = 1120, $\mu = 223.1$, $\sigma = 207.3$). The citation analysis on Scopus\footnote{We also considered Web of Science and Google Scholar databases, but disregarded them, since Web of Science does not index all years 2010--2019 of both conferences, and Google Scholar indexes a lot of lower-quality citations, such as bachelor's theses.} showed that the median citation count is 2 (min = 0, max = 18, $\mu = 3.3$, $\sigma = 4.4$). After removing self-citations, the median dropped to 1 (min = 0, max = 17, $\mu = 2.5$, $\sigma = 3.8$). We also looked at how many of the non-self-citations are from the SIGCSE or ITiCSE community. The median is 0 (min = 0, max = 5, $\mu = 0.5$, $\sigma = 1$). This shows that the papers are rarely cited, not even within the community. Nevertheless, these metrics are biased toward older papers, which have a higher chance of being downloaded or cited compared to the recently released papers. Lastly, to complement the point of view, the examined papers themselves include a median of 18 references (min = 4, max = 39, $\mu = 18.1$, $\sigma = 7$).

\subsection{RQ5: Who Forms the SIGCSE and ITiCSE Cybersecurity Community?}

Finally, we examined the people that publish cybersecurity education papers and their affiliations.

\subsubsection{Who publishes cybersecurity education research?}
Within the selected papers, there were 251 author listings and 202 unique authors\footnote{We manually double-checked the automatic analysis of authors' names to account for minor differences in how the same authors list their names in different papers, for example, Heather Richter Lipford~\cite{Tabassum:2018} vs. Heather Lipford-Richter~\cite{Whitney:2015}.}, out of which 175 -- a vast majority -- appeared only in one paper. This implies a lack of cybersecurity education researchers dedicated to publishing at SIGCSE and ITiCSE. However, 14 authors appeared in two papers, and the remaining 13 authors appeared in three or more papers, which suggests that there is a small but stable community of cybersecurity educators. The most prolific authors with five papers were Ching-Kuang Shene and Jean Mayo with their series of visualization papers on ITiCSE~\cite{Wang:2014, Wang:2015, Ma:2016, Wang:2017, Walker:2019}.

\subsubsection{What are the authors' affiliations?}
We looked at the authors' affiliations and countries to better understand the demographics of the community. Out of the 251 author listings, the vast majority were affiliated to universities (190) and colleges (31), following with military institutions (22), research centers (7), and private companies (2)\footnote{One author in~\cite{Egelman:2016} had two affiliations, therefore, the sum of the affiliations is 252.}. The most represented country was the USA (203), followed by Canada (17) and the Czech Republic (10). This corresponds to the fact that most SIGCSE/ITiCSE papers examine higher education interventions within the context of the USA.

\section{Discussion and Conclusions}

We performed a systematic literature review of 71 cybersecurity education papers from SIGCSE and ITiCSE, the leading conferences in the field of computing education, over the period from 2010 to 2019. Our dataset is publicly available as supplementary material in the ACM Digital Library and also on Zenodo~\cite{dataset}. Apart from reviewing the current literature, we also provided a framework for future researchers by listing the possible evaluation goals, evidence types, and analysis methods. We now summarize the most commonly observed trends in the examined papers and provide recommendations for both research and practice.

\subsection{Summary of the Observed Trends}

A typical SIGCSE/ITiCSE cybersecurity education paper deals with topics such as secure software development, network security, cyber attacks, cryptography, or privacy. It describes a course, hands-on exercise, or a tool applied in teaching practice, in the context of a North American university. It usually reports data and teaching experience from a period of one semester, with a population of several dozens of undergraduate students.

Considering the research goals, the typical evaluation examines subjective experiences and perceptions of students, using questionnaires as the most common research tool. Also, pre- and post-test study designs are standard to examine learning gains after the applied teaching intervention. The evaluation is performed on a subset of the student population. The results are presented with descriptive statistics; sometimes, inferential statistics are used to confirm relationships within the data.

Even though most papers mention creating new tools or teaching materials, only 31\% of the papers provide an output available to other educators and researchers. Surprisingly, no paper includes a dataset as supplementary material. Finally, a small number of citations of the papers may suggest that cybersecurity education research is fragmented. A possible explanation is that the researchers explore disjoint topics and rarely use others' results. What is more, almost 87\% of the unique authors who contributed to SIGCSE/ITiCSE cybersecurity education research did so only once.

\subsection{Implications of this Literature Review}

Several research ideas stem from this review. Since K-12 education was underrepresented, it may be worthwhile to examine teaching interventions with younger learners. Next, since not all students participated in the evaluation, exploring how to motivate them to take part in education research can be valuable. Moreover, as most papers used questionnaires or tests for evaluation, researchers may consider employing approaches of educational data mining or learning analytics to better understand students' learning processes. Lastly, it would be interesting to compare research trends in cybersecurity conferences with computing education conferences.

To support high-quality research, we recommend future authors to familiarize themselves with exemplary papers and subsequently perform more rigorous evaluations while sharing their datasets. The community would also benefit from more thorough reporting of the research methods. In some papers, the description of the methods was unclear or incomplete, complicating the extraction of data for this review. To support teaching practitioners, we suggest using standardized documents such as the JTF Curriculum~\cite{cybered} to precisely define learning outcomes and addressed topics. Also, using stable public repositories to share content would be beneficial, since the tools published in 8 out of 30 papers are no longer accessible.

\begin{acks}
This research was supported by the \grantsponsor{ERDF}{ERDF}{} project \textit{CyberSecurity, CyberCrime and Critical Information Infrastructures Center of Excellence} (No. \grantnum{ERDF}{CZ.02.1.01/0.0/0.0/16\_019/0000822}). We also thank Pavel Šmerk for his assistance with the citation analysis.
\end{acks}

\bibliographystyle{ACM-Reference-Format}
\bibliography{references}

\end{document}